\documentclass[iicol,sn-mathphys]{sn-jnl}
\usepackage{bm,graphicx}

\newcommand{\beq}{\begin{equation}}
\newcommand{\eeq}{\end{equation}}
\newcommand{\beqa}{\begin{eqnarray}}
\newcommand{\eeqa}{\end{eqnarray}}

\newcommand{\abs}[1]{\vert#1\vert}
\newcommand{\dd}{{\rm d}}
\renewcommand{\e}{{\rm e}}
\newcommand{\eff}{{\rm eff}}
\newcommand{\erfc}{\mathop{\rm erfc}}
\newcommand{\frad}[2]{\frac{\displaystyle#1}{\displaystyle#2}}
\renewcommand{\max}{{\rm max}}
\newcommand{\mean}[1]{\langle#1\rangle}
\newcommand{\para}{{\rm para}}
\newcommand{\prob}[1]{{\rm Prob}(#1)}
\renewcommand{\var}{\mathop{\rm var}}
\renewcommand{\vec}[1]{{\bm#1}}
\newcommand{\win}{{\rm win}}

\begin{document}

\title{Evolution of grammatical forms: some quantitative approaches}

\author*[1]{\fnm{Jean-Marc} \sur{Luck}}\email{jean-marc.luck@ipht.fr}
\author[2]{\fnm{Anita} \sur{Mehta}}\email{anita.mehta@ling-phil.ox.ac.uk}

\affil[1]{\orgname{Universit\'e Paris-Saclay, CNRS \& CEA},
\orgdiv{Institut de Physique Th\'eorique},
\city{91191~Gif-sur-Yvette}, \country{France}}

\affil[2]{\orgdiv{Faculty of Linguistics, Philology and Phonetics},
\orgname{Clarendon Institute, Walton Street},
\city{Oxford OX1 2HG}, \country{United Kingdom}}

\abstract{Grammatical forms are said to evolve via two main mechanisms.
These are, respectively, the `descent' mechanism, where current
forms can be seen to have descended (albeit with occasional modifications) from
their roots in ancient languages, and the `contact' mechanism, where
evolution in a given language occurs via borrowing from other languages with
which it is in contact.
We use ideas and concepts from statistical physics
to formulate a series of static and dynamical models which illustrate these
issues in general terms.
The static models emphasise
the relative numbers of rules and exceptions, while the dynamical models focus
on the emergence of exceptional forms.
These unlikely survivors among various competing grammatical forms
are winners against the odds.
Our analysis suggests that they emerge
when the influence of neighbouring languages
exceeds the generic tendency towards regularisation within individual languages.}

\maketitle

\section{Introduction}
\label{intro}

Historical linguistics is the study of language change over time~\cite{bynon}.
It is principally centred on how linguistic forms evolve in world languages,
be these to do with phonology, morphology, semantics, syntax or core lexicons.
The evolution of languages proceeds in two essentially different ways.
The first route, referred to as phylogeny,
describes the `vertical' descent with modification from more ancient,
possibly extinct languages,
leading to the representation of families of languages as branching
phylogenetic trees.
The second comprises the `horizontal' borrowing and diffusion between
contemporary languages,
brought about via contact between their speakers; this has given rise to the
field of contact linguistics~\cite{TK,winford}.

The mechanisms involved during language contact will be of special
importance in this paper, so we introduce them here.
Winford~\cite{winford} has classified contact
between different linguistic communities under three different heads: he describes, first,
relatively homogeneous communities of monolinguals, most of whom have little
or no contact with speakers of other languages.
In these, the only way that borrowing
occurs is via the media, or individual travellers, or indeed foreign language
teaching; the example of Japanese or Russian speakers borrowing from English is
cited as an example.
The so-called `middle spectrum' concerns communities which include bilingual or
multilingual speakers, an example of which might be the contact between
linguistic minorities and a dominant host group; the prevalence of French and
Flemish in Brussels is cited as an example.
Finally, there are highly heterogeneous communities
where individual multilingualism is high and social and linguistic boundaries
are fluid; an example of this occurs in Northwest New Britain in Papua New Guinea.
The nature of the mutual influence between different
languages depends critically on the extent of contact~\cite{winford}:
slight lexical borrowing
is about all that happens under conditions of casual contact, while
structural borrowing needs the concerned linguistic communities to undergo
sustained and intimate contact.
In particular this implies that while the borrowing of
content morphemes like nouns or verbs is common, the borrowing of
grammatical features is relatively rare.

There has been relatively little work done so far from the viewpoint of statistical physics
in modelling linguistic evolution,\footnote{In contrast,
such concepts have been used in other areas of language dynamics,
such as the coexistence of two or more languages in a given geographical area~\cite{bly}.}
although several quantitative approaches~\cite{ross,warnow,lieberman,peter1,pnas,peter2,jacques}
exist.
In this work, we apply statistical physics methodologies
to explore the evolution of morphological features in general, and rules governing
verb conjugation in particular.
Our initial motivation for this was a question raised by Ringe and Yang~\cite{longringe}
concerning the evolution of past participles of verbs in English, via
the `tolerance principle'~\cite{yang}; this states that
that there is a maximum number of exceptions that a rule can tolerate in order to be productive.
More precisely, a rule applied to $N$ items
obeys the tolerance principle if the number~$E$ of exceptions
(i.e., of items to which it does {\it not} apply)
is smaller than the threshold
\beq
E_N=\frac{N}{H_N}\approx\frac{N}{\ln N}
\label{tole}
\eeq
($H_N$ is the $N$th harmonic number).
From the viewpoint of statistical physics, this threshold $E_N$ is enormously high,
since it grows nearly extensively with the total number $N$ of items.
Our own evaluations of a threshold demarcating rules and exceptions
(Section~\ref{static}) result in
smaller and more realistic values, to which we will draw attention as they occur.

Ringe and Yang claimed that the tolerance principle was able to
explain the prevalence of regular past participles ending in `-ed', but not
some more unlikely irregular forms ending in `-uck', such
as `stuck' or `struck'~\cite{longringe}.
Our models of competitive dynamics are able
to resolve this issue, while also putting the emergence of unlikely winners in
a more general context.

The very formulation of the tolerance principle gives a prominent role to exceptions.
Grammatical rules are an essential feature of linguistic structure
and provide an efficient way of classifying existing forms;
there are, however, always exceptions to these.
For instance, the conjugation of the verbs
`to be' or `to have' is rather irregular in most world languages, so that these verbs
constitute marked exceptions to general grammatical rules.

The occurrence of rules and exceptions is not unique to languages.
In mathematics, for example,
the classification of semisimple Lie algebras involves~4 rules
(the infinite series $A_n$, $B_n$,~$C_n$, and~$D_n$)
and~5 exceptions ($E_6$, $E_7$, $E_8$, $F_4$ and~$G_2$).
Also, the classification of finite simple groups involves 18 rules (infinite families of groups)
and 26 exceptions (the sporadic groups) (see~\cite{solomon}).
Note also that the numbers of rules and exceptions
(4 vs.~5 and 18 vs.~26) are comparable in the two cases.
Rules and exceptions also occur naturally in data clustering,
in the context of computer science and data analysis, where
they are termed clusters and outliers respectively (see e.g.~\cite{BL,out}).
Clearly, in all of the above, rules concern either large or infinite series of objects, while
exceptions are isolated.

The plan of this paper is as follows.
We begin with a purely static approach to the interplay
between grammar rules and exceptions (Section~\ref{static}),
which in particular results in sensible values for the threshold that divides them.
In the following sections,
we build increasingly sophisticated models of the dynamical evolution of grammar rules.
Successive levels of modelling include the initial growth of the structured lexicon
in a single language (Section~\ref{growth}),
the competition between growth and conversions, e.g.~from irregular to regular forms,
in a mature language (Section~\ref{gandc}),
and finally a network representation of the comparative evolution of grammar rules
in a situation of prolonged language contact (Section~\ref{comparative}).
Finally, we summarise and collate our insights in the Discussion section
(Section~\ref{disc}).

\section{Rules and exceptions: a~static approach}
\label{static}

This section provides a first approach to the interplay between grammar rules and exceptions.
The arguments used come from a purely static perspective, being
based on optimality: no dynamical evolution is invoked.

We focus here on morphology, and in particular on rules governing verb conjugation.
Consider a language having a total of $N$ verbs,
which are divided into $K$ groups.
Each group follows a distinct pattern of conjugation,
and is labelled by an integer $a=1,2,\dots,K$.
Group $a$ contains $N_a$ verbs, so that
\beq
N=\sum_{a=1}^KN_a.
\eeq
Groups are assumed to be ranked in order of decreasing size ($N_1\ge N_2\ge N_3\dots$).

It is clearly most efficient to remember the conjugations of the $N_1$ verbs in
the largest group by means of a single rule.
On the other hand, if the smallest groups ($a$ near $K$) have $N_a=1$,
it is most efficient to think of them as exceptions
(e.g.~verbs such as `to be' or `to have' in most languages).
How, then, can a demarcation line between rules and exceptions be operationally defined?
A natural way of proceeding consists of minimising
the total memorization effort and memory size $I$ needed to learn and store
the conjugations of all verbs comprising both rules and exceptions,
or, in Chomsky's~\cite{chomsky} words, `the grammar' and `the lexicon'.
If the conjugation in question has $R$ rules,
the latter correspond to the~$R$ largest groups.
The remaining $K-R$ groups comprise a total of~$E$ exceptions, with
\beq
E=\sum_{a=R+1}^KN_a.
\label{edef}
\eeq
We estimate the requested memory size as
\beq
I=CR+E,
\label{idef}
\eeq
where $C$ is the only free parameter of the model.
It obeys the inequality $C>1$,
expressing our expectation that it takes more effort and memory size
to remember a full rule than to remember an exception.
In particular, the above inequality
ensures that single verbs, belonging to groups with $N_a=1$,
are automatically considered as exceptions.
Minimising $I$ with respect to $R$ should yield the optimal number of rules.

\subsection{Exponential size distribution}

Consider first the situation where group sizes have an exponential asymptotic
decay of the form
\beq
N_a\approx AN\e^{-\mu a},
\label{naexp}
\eeq
for some constants $A$ and $\mu$.
Since group sizes are obviously integers, we need to take the integer part
of the right-hand side of~(\ref{naexp}).
However, neglecting this subtlety, we get accurate asymptotic estimates
for the quantities of interest
in the realistic situation where the total number $N$ of verbs is large,
whereas the parameter $\mu$ is small.
Setting $N_K=1$ yields an estimate for the total number of groups,
\beq
K\approx\frac{1}{\mu}\ln AN.
\eeq
For a given number $R$ of rules,~(\ref{edef}) yields
\beq
E\approx\frac{AN}{\e^\mu-1}\e^{-\mu R}.
\eeq
The total memory size $I$ is minimal for
\beq
R\approx\frac{1}{\mu}\ln\frac{\mu AN}{C(\e^\mu-1)}.
\label{log1}
\eeq

This approach predicts that the number of rules grows logarithmically
with the total number of verbs.
Considering the relatively small number of conjugation rules
in most world languages, this slow growth of the number of rules makes very good sense.
Our approach also predicts that the number $E$ of exceptions saturates to the finite limit
\beq
E\approx\frac{C}{\mu}.
\eeq
The actual integer value of $E$ oscillates around the above limit, which, we note,
is much smaller than the threshold~(\ref{tole}) predicted by the tolerance principle.

\subsection{Power-law size distribution}

If group sizes have a power-law decay, namely
\beq
N_a\approx\frac{BN}{a^{\theta+1}},
\eeq
with an arbitrary positive exponent $\theta$,
we have
\beq
K\approx(BN)^{1/(\theta+1)}.
\eeq
For a given number $R$ of rules,~(\ref{edef}) yields
\beq
E\approx\frac{BN}{\theta R^\theta}.
\eeq
The memory size $I$ is minimal for
\beq
R\approx\left(\frac{BN}{C}\right)^{1/(\theta+1)}.
\eeq
We have then
\beq
E\approx\frac{C}{\theta}\,R.
\eeq

In this case, the numbers $R$ of rules and $E$ of exceptions grow proportionally to each other.
Their common growth law is subextensive in the total number $N$ of verbs,
and characterised by the growth exponent $1/(\theta+1)$.
This prediction for~$E$ is again much smaller than the threshold~(\ref{tole})
predicted by the tolerance principle.

\section{A dynamical model with growth}
\label{growth}

This section contains the first of several dynamical approaches
to the evolution of grammar rules and exceptions.
We adopt a chronological viewpoint which assumes that new verbs are added
sequentially to the lexicon.
Each new verb typically joins an existing group and follows its conjugation rules,
whereas it rarely, if ever, forms a new group.

This model is freely inspired from
the theory of growing networks by preferential attachment,
proposed by Barabasi and Albert~\cite{ba1,ba2}.
Among various extensions of the model~\cite{doro,KRR},
Bianconi and Barabasi~\cite{bb1,bb2} have shown that
the addition of a fitness or attractiveness parameter characterising each node
greatly enriches the model; among other things, it
may induce a condensation phase transition.

Thus, new verbs enter the lexicon sequentially as new nodes in growing networks.
At any given instant, there are $K$ verb groups, indexed $a=1,\dots,K$.
Group $a$ contains $N_a$ verbs subject to specific grammar rules governing
their conjugation.
The total number of verbs then reads\footnote{This number will be used
as an effective measure of `time'.
In this work, we never directly compare real (i.e., historical) time
to the effective time variables parametrising the evolution in all our models.}
\beq
N=\sum_{a=1}^KN_a.
\eeq

In this approach, exceptions are not considered explicitly,
so that the number $K$ of groups is identical to the number $R$ of rules.

The new verb number $(N+1)$ joins group $a$ with probability
\beq
p_a=\frac{\eta_a(N_a+c)}{Z(N)},
\eeq
where:

\begin{itemize}

\item
The first factor $\eta_a$ is the intrinsic attractiveness (or fitness)
parameter of group $a$.
It is fixed once and for all at the birth of group $a$,
and embodies the Darwinian {\it fit-get-richer} effect.

\item
The second factor $(N_a+c)$ is dynamical, in the sense that it grows in the course of time.
It embodies the {\it rich-get-richer}, or Matthew, effect.
The constant $c$ is the initial attractiveness of an empty
group~\cite{doro,KRR}.

\item
The denominator
\beq
Z(N)=\sum_a\eta_a(N_a+c)
\eeq
ensures the normalisation of the attachment probabilities $p_a$ at all times.

\end{itemize}

\subsection{Evolution of number of groups}

A new verb group is born, i.e., $K$ is changed to $K+1$,
whenever the new verb starts it, instead of joining an existing one.
The geometric picture behind this is the growth of a forest of $K$ trees,
where the joining of a new verb to an existing group $a$
corresponds to the growth of the $a$th tree,
whereas a newborn tree appears whenever the incoming verb itself starts a new group.

The birth of a new group takes place with probability $c\eta_{K+1}/Z(N)$.
In the late stages of the dynamics, when group sizes are large,
the following deterministic growth law emerges
\beq
\frac{\dd K}{\dd N}\approx\frac{c}{N},
\eeq
yielding
\beq
K\approx c\ln N.
\label{klog}
\eeq
The logarithmic growth laws~(\ref{log1}) and~(\ref{klog}) substantiate our intuition
that the birth of a new grammar rule is a rare event.

\subsection{Evolution of group sizes}
\label{gpdyn}

Consider again the late stages, where group sizes are typically large.
For a given draw of the attractiveness parameters $\eta_a$ of the groups,
the stochastic rules defining our growth model reduce
asymptotically to the deterministic growth equations
\beq
\frac{\dd N_a}{\dd N}\approx\frac{\eta_aN_a}{Z(N)},
\label{geqs}
\eeq
with
\beq
Z(N)\approx\sum_a\eta_aN_a.
\eeq

Discarding the rare events where a new verb group is born,
so that the number $K$ of groups remains constant,
we rank verb groups according to decreasing attractiveness
($\eta_1>\eta_2>\dots>\eta_K$).
The size of the most attractive group grows as $N_1\approx N$.
We have therefore $Z(N)\approx\eta_1N$,
so that the sizes of the other groups ($a=2,\dots,K$) asymptotically obey
\beq
\frac{\dd N_a}{\dd N}\approx\frac{\eta_aN_a}{\eta_1N},
\eeq
and hence
\beq
N_a\sim N^{\beta_a},
\label{nexp}
\eeq
with
\beq
\beta_a=\frac{\eta_a}{\eta_1}.
\eeq
Our prediction is that, apart from the most favoured one,
group sizes grow with a subextensive power-law,
the growth exponents $\beta_a<1$ being given by attractiveness ratios.
A similar power-law growth scenario with variable exponents
holds in the Bianconi and Barabasi model of a growing network~\cite{bb1,bb2}.

This subextensive growth law~(\ref{nexp}) with continuously variable exponents $\beta_a$
suggests a smooth crossover between rules and exceptions,
instead of a sharp line of demarcation dividing the two
(see the static approach of Section~\ref{static}).
Notice once again that the predicted sizes of all unfavoured groups
are much smaller than the threshold~(\ref{tole}) involved in the tolerance principle.

\section{A dynamical model with growth and conversions}
\label{gandc}

This second dynamical approach describes a later stage in the evolution of a mature language.
Verbs might spontaneously change groups,
converting, for instance, from an irregular to a regular form.
This conversion mechanism competes with the growth mechanism introduced
in Section~\ref{growth}; it may result in the enrichment and eventual dominance of
a verb group which is not {\it per se} the most attractive.
This is a manifestation of the phenomenon of winning {\it against the odds},
which we have explored in various contexts~\cite{us1,us2,us3}.

Again, we discard the rare events where a new group of
verbs is born, so that the number~$K$ of groups is constant,
and the group label~$a$ runs from~1 to~$K$.
The total rate of conversions from group~$b$ to group~$a$ is assumed
to be proportional to the sizes of both groups.
It therefore reads $C_{ab}N_aN_b$,
where the individual conversion rates~$C_{ab}$ are the entries
of a constant skew-symmetric conversion matrix of size $K\times K$.

In the presence of conversions, the evolution equations~(\ref{geqs}) therefore
read
\beq
\frac{\dd N_a}{\dd
N}=N_a\left(\frac{\eta_a}{Z(N)}+\frac{1}{N^2}\sum_bC_{ab}N_b\right),
\label{gceqs}
\eeq
with
\beq
N=\sum_aN_a,\quad Z(N)=\sum_a\eta_aN_a.
\eeq
The first and second terms in the parentheses in~(\ref{gceqs})
respectively describe the competing growth and conversion mechanisms.
The second term has been rescaled by $1/N^2$,
in order to ensure that the strengths of both competing mechanisms
remain comparable in the regime of large $N$, i.e., for very mature languages.
The description of this conversion mechanism is similar to that used
in earlier work on the coexistence of two or more languages in competition~\cite{coex}.

The relative sizes of the various verb groups, defined as the ratios
\beq
x_a=\frac{N_a}{N}\quad(a=1,\dots,K),
\eeq
obey the reduced evolution equations
\beq
N\frac{\dd x_a}{\dd N}=x_a\left(\frac{\eta_a}{z}-1+\sum_bC_{ab}x_b\right),
\label{xeqs}
\eeq
with the sum rules
\beq
\sum_ax_a=1,\quad\sum_a\eta_ax_a=z.
\eeq

The dynamical system~(\ref{xeqs}) shares a remarkable property
with that encountered in earlier work~\cite{coex}.
For generic values of the attractiveness parameters $\eta_a$
and the conversion rates $C_{ab}$,
the coupled evolution equations~(\ref{xeqs}) have a {\it single} attractor,
which consists of an attractive fixed point $\vec{x^\star}=\{x_a^\star\}$.
In general, some components $x_a^\star$ are positive, whereas others vanish.

\begin{itemize}

\item
If $x_a^\star>0$, group $a$ is said to be a survivor.
Its size grows asymptotically in proportion to the total number of verbs, as
\beq
N_a\approx Nx_a.
\eeq

\item
If $x_a^\star=0$, group $a$ is said to be extinct.
According to the deterministic equations~(\ref{xeqs}),~$x_a$ falls off exponentially in $N$.
In the microscopic stochastic context,
this means that the group concerned goes extinct in a finite time.

\end{itemize}

The unique attractor of the dynamical system~(\ref{xeqs}) defines the pattern of survivors.
Its principal characteristic is the number $M$ of survivors.
Some specific sets of model parameters can be chosen so as to have either $M=1$
(corresponding to the emergence of a grammatical consensus,
in the sense that only one group survives), or $M=K$
(all groups survive), or an arbitrary number of groups
in the range $M=1,\dots,K$, survive.
We emphasise that in the present setting, the number $M$ of surviving groups is
identical to the number $R$ of rules.

From here on, we consider the specific situation where verbs
convert from less to more regular forms.
This is motivated by the observed general tendency for irregular verbs in most languages
to `regularise' with time (see e.g.~\cite{lieberman}),
even when the former are well established (see, for example, the gradual
regularisation of the past participle `wed'
to an increasingly accepted `wedded'~\cite{lieberman}).

For simplicity, we now rank verb groups according to decreasing regularity,
$a=1$ being the most regular, and $a=K$ the most irregular group.
The conversion mechanism then takes the form of a simple descent at some constant rate $g$.
Within this framework, the entries of the conversion matrix read
\beq
C_{ab}=\left\{
\begin{matrix}
+g\quad & (a<b),\cr
0\hfill & (a=b),\cr
-g\hfill & (a>b).
\end{matrix}
\right.%}
\eeq

\subsection{The case of two verb groups in competition}

We consider first the case of two verb groups
with attractiveness parameters $\eta_1$ and $\eta_2$.
For $K=2$, the reduced evolution equations~(\ref{xeqs}) read
\beqa
N\frac{\dd x_1}{\dd N}=x_1\left(\frac{\eta_1}{z}-1+gx_2\right),
\nonumber\\
N\frac{\dd x_2}{\dd N}=x_2\left(\frac{\eta_2}{z}-1-gx_1\right),
\label{x12eqs}
\eeqa
with
\beq
x_1+x_2=1,\quad\eta_1x_1+\eta_2x_2=z.
\eeq
The two control parameters are the attractiveness ratio
\beq
q=\frac{\eta_1}{\eta_2}
\eeq
and the conversion rate $g$.

\begin{figure}[!htbp]
\begin{center}
\includegraphics[angle=0,width=.8\linewidth,clip=true]{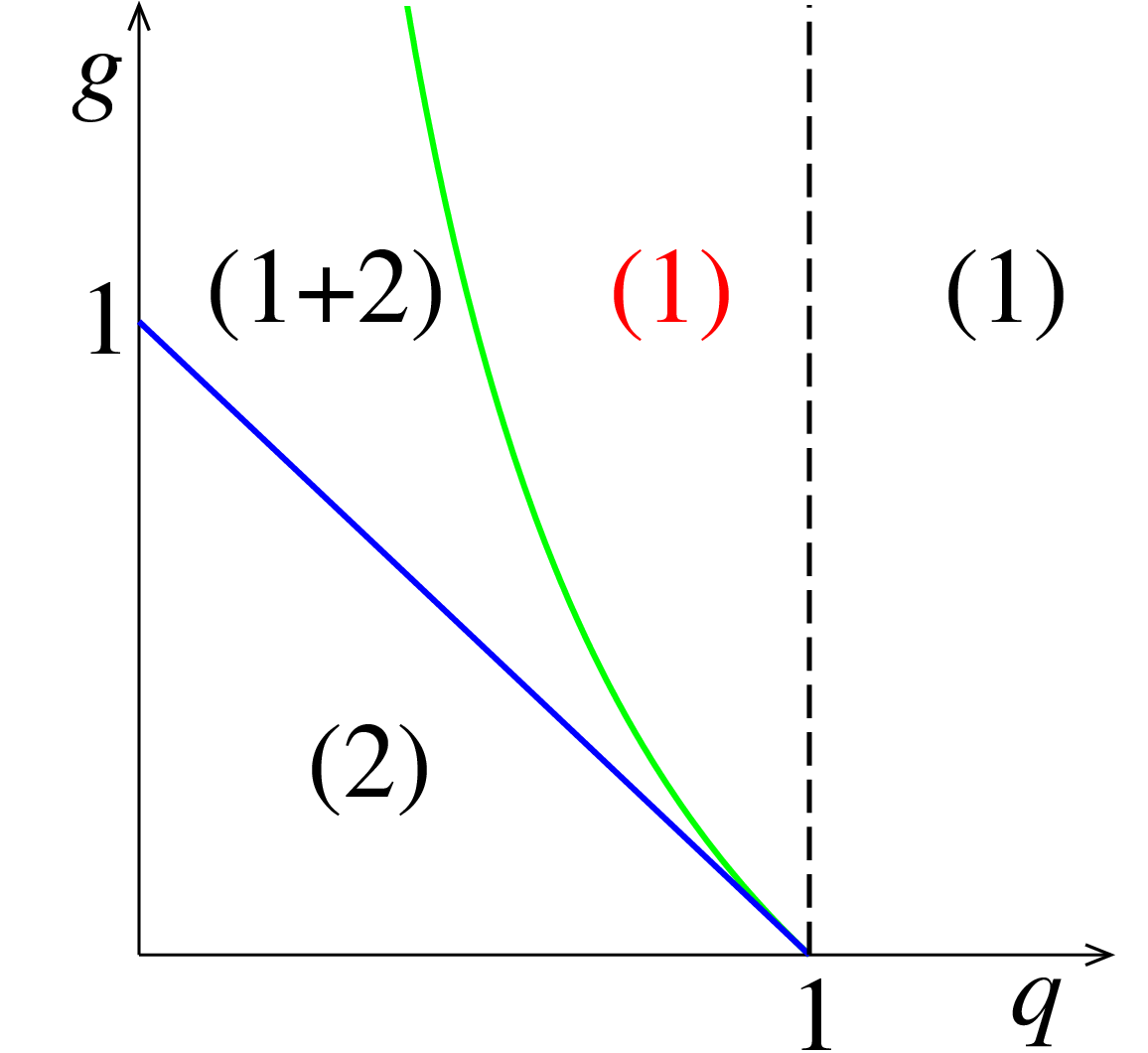}
\caption{
Phase diagram of the model of two verb groups
with growth and conversions in the $q$--$g$ plane.
Blue and green curves: boundaries of coexistence range.
Vertical dashed line: neutral line where $\eta_1=\eta_2$.
Numbers between parentheses: labels of surviving group(s).
Red label: survivor against the odds.}
\label{qgplot}
\end{center}
\end{figure}

Figure~\ref{qgplot} shows the phase diagram of the model in the $q$--$g$ plane.
In the absence of conversions $(g=0$),
only the most attractive verb group survives,
namely group~1 for $q>1$ and group~2 for $q<1$.
The presence of conversions triggers several novel phenomena.
First, both groups survive simultaneously in an intermediate range
\beq
1-q<g<\frac{1-q}{q},
\eeq
whose boundaries are respectively shown in blue and green.
In this coexistence range we have
\beq
x_1^\star=\frac{g+q-1}{(1-q)g},\quad
x_2^\star=\frac{1-q(1+g)}{(1-q)g}.
\eeq
The first fraction $x_1^\star$ (resp.~the second fraction~$x_2^\star$)
vanishes continuously as the blue (resp.~green) boundary is approached.
Second, the competition between the growth and conversion mechanisms
allows for the emergence of survivors against the~odds; thus,
in the domain between the green curve and the vertical dashed line (labelled in red),
group 1 survives despite being unfavoured.

\subsection{The general case of $K$ verb groups}

For a higher number $K\ge3$ of competing verb groups,
the determination of the full multidimensional phase diagram of the model is intractable.

We consider instead a statistical ensemble,
where the intrinsic attractiveness parameters $\eta_a$
of verb groups are modelled as independent quenched random variables
drawn from some probability distribution.
We choose for definiteness the exponential distribution with unit width:
\beq
f(\eta)=\e^{-\eta}.
\label{fexp}
\eeq
The main features of the model,
including the logarithmic growth law~(\ref{mlog}),
would remain qualitatively unchanged
for any bounded or rapidly decaying attractiveness probability distribution.

The survivors thus form a random pattern,
whose statistics depend only on the number $K$ of groups
and the conversion rate $g$.
This pattern can be easily predicted at small and large $g$.
If $g$ is either zero or very small,
the growth mechanism dominates and only the most attractive group survives.
If $g$ is very large, the conversion mechanism wins; now the only survivor is
the most regular group.
In the intermediate regime where the conversion rate $g$
is moderate, so that growth and conversion are comparable, several survivors may coexist.

\begin{figure}[!htbp]
\begin{center}
\includegraphics[angle=0,width=1\linewidth,clip=true]{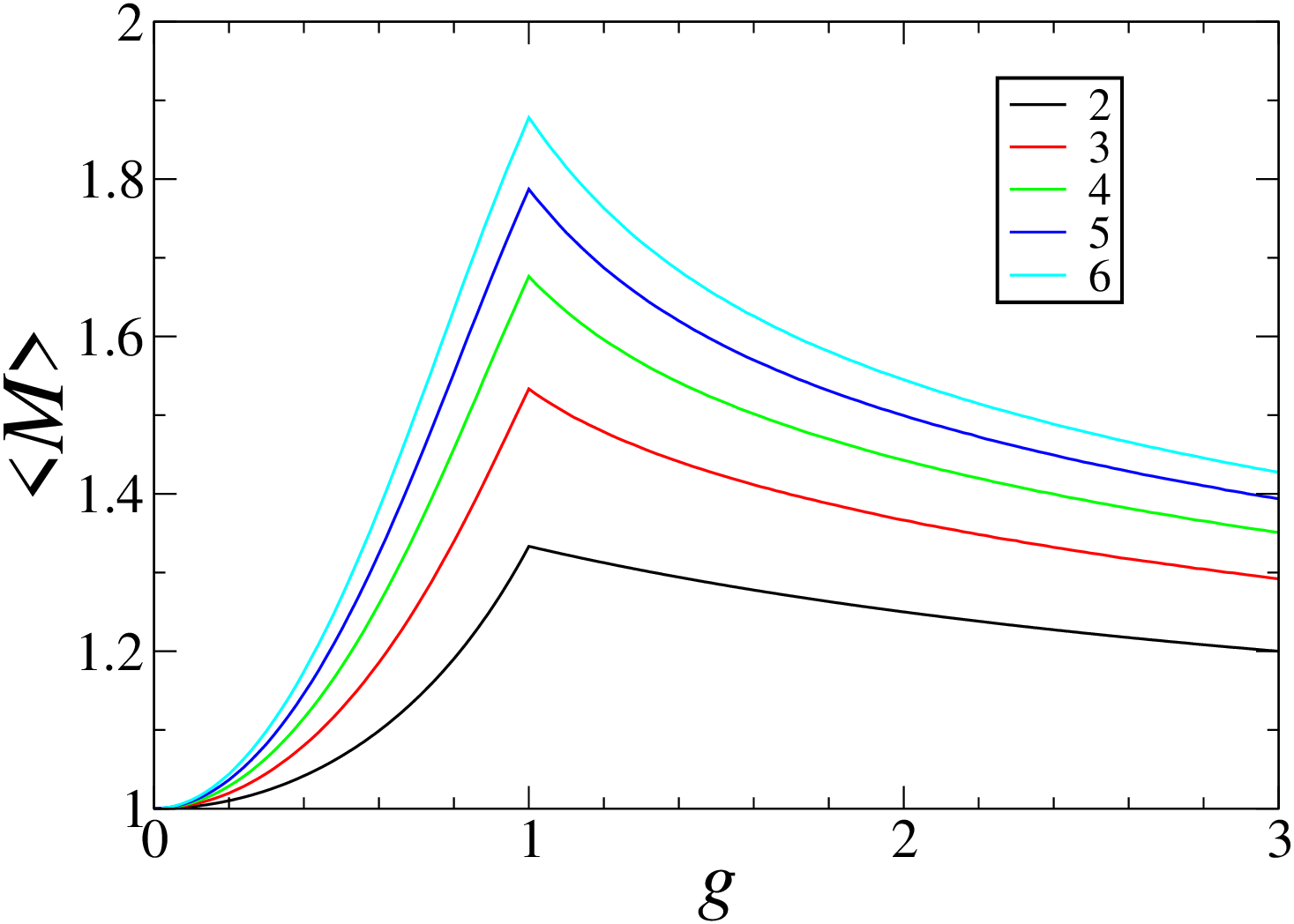}
\caption{
Mean number $\mean{M}$ of survivors against conversion rate $g$
for several numbers $K$ of verb groups (see legend).
Black curve: exact analytical result~(\ref{mave2}) for $K=2$.
Coloured curves: numerical data for $K=3$ to 6.}
\label{mave}
\end{center}
\end{figure}

We first focus on the mean number $\mean{M}$ of survivors,
where the mean value is taken over the distribution~(\ref{fexp}) of
attractiveness parameters.
In the case of two verb groups ($K=2$),
the exactly known phase diagram of the model (see Figure~\ref{qgplot}) yields
\beq
\mean{M}=\left\{
\begin{matrix}
\frad{4}{4-g^2}\quad & (g\le1),\cr\cr
\frad{g+3}{g+2}\quad & (g\ge1).
\end{matrix}
\right.%}
\label{mave2}
\eeq
This expression goes to unity at small and large~$g$, as expected.
It takes its maximal value, $\mean{M}=4/3$, at $g=1$; its
first derivative is discontinuous at this point, as indicated by the cusp in
the black curve in Figure~\ref{mave}.
This singularity is due to the endpoint at $(q=0,g=1)$
of the blue line in Figure~\ref{qgplot}.

We have investigated the behaviour of larger systems ($K\ge3$)
by means of numerical simulations,
determining the value of the fixed point
$\vec{x^\star}=\{x_a^\star\}$
for many independent draws ($10^7$ for each parameter set) of the
attractiveness parameters $\eta_a$.
Figure~\ref{mave} shows the mean number $\mean{M}$ of survivors
plotted against the conversion rate $g$ for several $K$ (see legend).
The black curve shows the exact analytical result~(\ref{mave2}) for $K=2$.
The other curves show the outcome of numerical simulations.
The qualitative form of the dependence of~$\mean{M}$ on $g$ is independent of $K$,
with $\mean{M}$ departing quadratically from unity at small $g$,
reaching its maximum with a cusp at $g=1$,
and slowly converging back to unity at large $g$.

\begin{figure}[!htbp]
\begin{center}
\includegraphics[angle=0,width=1\linewidth,clip=true]{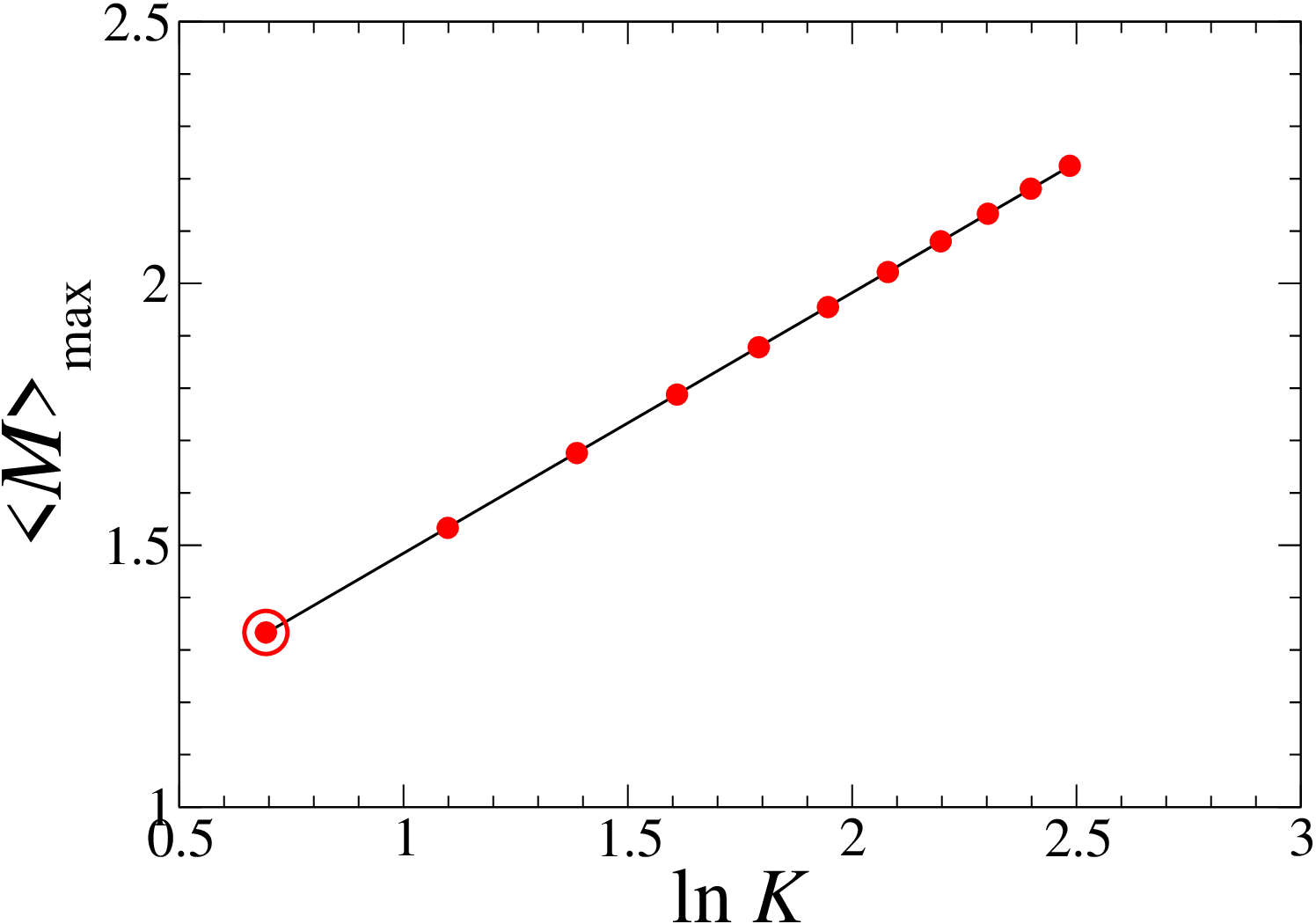}
\caption{
Maximal mean number $\mean{M}_\max$ of survivors,
corresponding to $g=1$, plotted against $\ln K$, for a number $K$ of verb groups up to 12.
Symbols: numerical data.
Circled symbol: exact value $\mean{M}_\max=4/3$ for $K=2$.
Full line: least-square fit with slope 0.498.}
\label{mmax}
\end{center}
\end{figure}

Figure~\ref{mmax} shows the maximal mean number $\mean{M}_\max$ of survivors,
corresponding to the conversion rate $g=1$, plotted against $\ln K$, for $K$ up to 12.
The excellent agreement with the regression line demonstrates that this quantity
grows logarithmically with the number $K$ of groups, as
\beq
\mean{M}_\max\approx A\ln K,
\label{mlog}
\eeq
with a prefactor $A\approx0.5$.
The appearance of this logarithmic law again emphasises the conformity of our model
with the general principles laid out in earlier sections (see~(\ref{log1}),~(\ref{klog})).
In the present context, the clear implication of this law is that
only very few verb groups survive from an initial panoply of possibilities.

In addition to the number $M$ of survivors, the whole pattern of survivors is also of interest.
In the absence of conversions, only the most attractive group survives.
As the conversion rate $g$ increases,
survivors against the odds~\cite{us1,us2,us3} -- i.e., those which do not
belong to the most attractive groups --
gradually become more and more frequent.
We define the variable $P_\para$, the `paradoxical' probability
that the most attractive group does {\it not} belong to the pattern of survivors,
to explore this issue further.

In the case of two verb groups ($K=2$),
the paradoxical probability $P_\para$ is nothing but the statistical weight
of the region lying between the green curve and the vertical dashed line
in the phase diagram of the model (see Figure~\ref{qgplot}), which is evaluated as
\beq
P_\para=\frac{g}{2g+4}.
\label{pro2}
\eeq
The relevant region does not touch the endpoint $(q=0,g=1$).
Hence, and at variance with the mean number of survivors,
$P_\para$ has a smooth dependence on $g$.

\begin{figure}[!htbp]
\begin{center}
\includegraphics[angle=0,width=1\linewidth,clip=true]{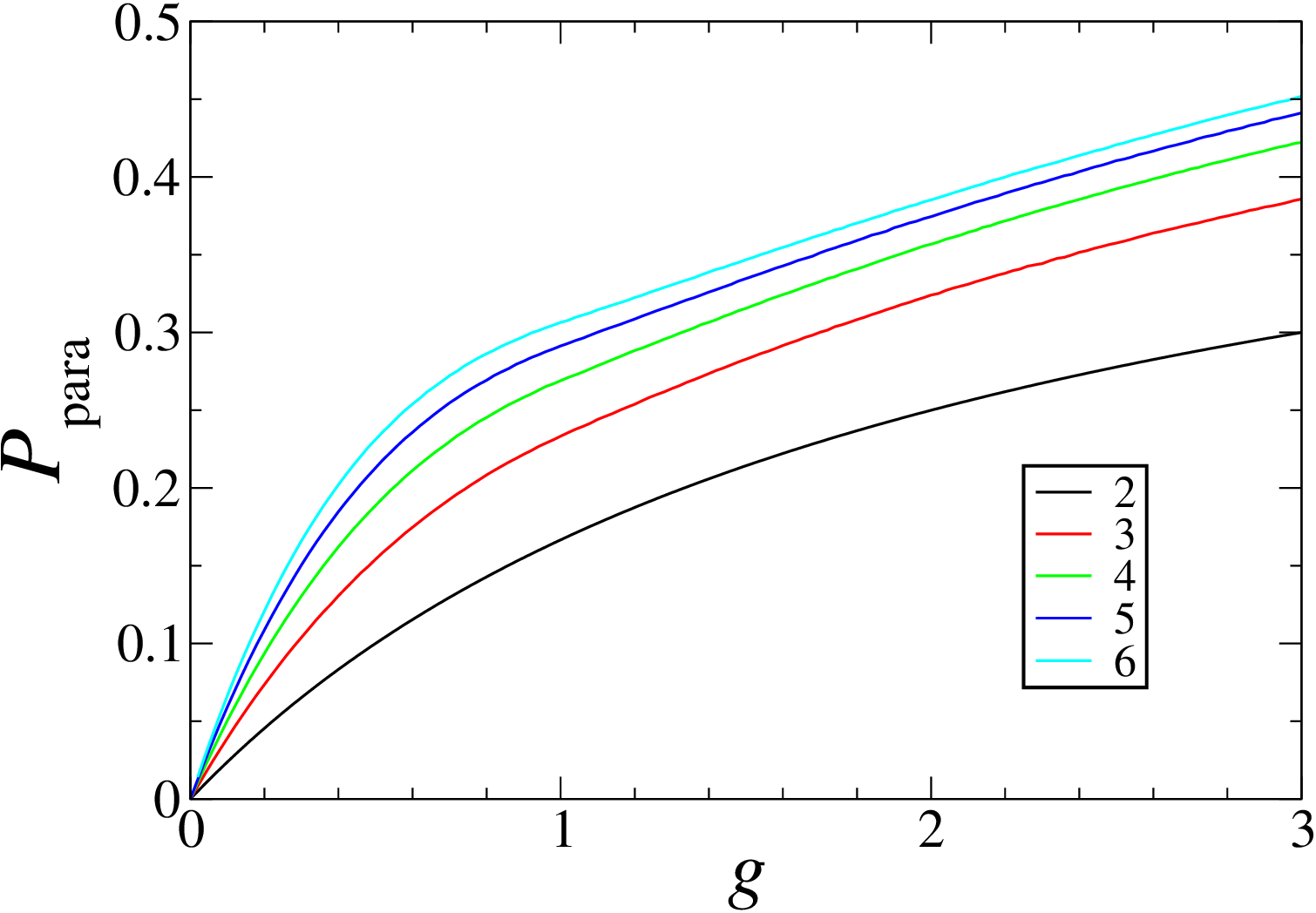}
\caption{
Paradoxical probability $P_\para$ against conversion rate $g$
for several numbers $K$ of verb groups (see legend).
Black curve: exact analytical result~(\ref{pro2}) for $K=2$.
Coloured curves: numerical data for $K=3$ to 6.}
\label{pro}
\end{center}
\end{figure}

Figure~\ref{pro} shows the paradoxical probability~$P_\para$
plotted against the conversion rate $g$ for several values of $K$ (see legend).
The black curve shows the exact analytical result~(\ref{pro2}) for $K=2$.
The other curves show the outcome of numerical simulations.
As expected, $P_\para$ increases steadily as a function of the conversion rate $g$,
departing linearly from zero at $g=0$,
exhibiting a shoulder for $g$ slightly below unity,
and slowly saturating to the limit value
\beq
P_\infty=\frac{K-1}{K}
\eeq
at very large $g$.
For infinitely large $g$, there is indeed only one survivor,
namely the most regular group ($a=1$), whose probability of also being
the most attractive one is $1/K$.

\section{Languages in contact}
\label{comparative}

In the previous section, we showed that conversions provided a mechanism for
gradual regularisation of initially irregular grammatical forms.
This occurs with increasing usage within the same language~\cite{lieberman}, i.e.
it is an intra-language mechanism.

On the other hand, our initial motivation for this study was the unlikely
survival of {\it irregular} past participles
such as `stuck'', when the favourite to win at the time was clearly `sticked'~\cite{longringe}.
We suggest that this emergence is due to contact with
other languages, i.e., that it is attributable to an inter-language mechanism.

As mentioned in the Introduction,
the field of contact linguistics concerns such linguistic influence.
We cite below a couple of instances of languages in prolonged contact,
which have led to their deep modification~\cite{winford}.
The first concerns the contact of
Old English with Norse, and then that of Middle English with Norman French,
which were the precursors of English in its present form.
The second concerns the Balkan {\it Sprachbund},\footnote{This term, meaning
`union of languages' refers to a situation where there is prolonged
contact across geographically contiguous language communities~\cite{winford}.}
where speakers of Greek, Romanian, and various Slavic languages were
in contact for almost a millennium.
In both cases, there has been an appreciable amount
of convergence in the morphology and syntax of the languages in contact,
despite the sizeable differences between them originally.

In the following, we model the effect of such contact among a given family of languages.
We represent all verb groups of this linguistic family as the nodes of a
graph, and the couplings between them as bonds connecting these nodes.
Bonds which connect nodes pertaining to the same language
represent the conversion mechanism introduced in Section~\ref{gandc},
whereas those joining nodes pertaining to different languages
represent the new ingredient of linguistic contact.
This description of language contacts is static,
in the sense that the topology of the network does not change over the course of time.
It therefore describes e.g.~the influence of Norse on Old English,
or that of French on Middle English, but not both together.

For the sake of simplicity, we write the corresponding evolution equations
in the following linear form:
\beq
\frac{\dd N_a}{\dd N}=\frac{\eta_a}{Z(N)}\left(N_a+\sum_bg_{ab}N_b\right).
\label{ceqs}
\eeq
These equations present analogies and differences with the evolution equations~(\ref{gceqs}).
In both cases~$\eta_a$ is the intrinsic attractiveness parameter of group $a$.
Within the present linear setting,
the couplings $g_{ab}$ represent the strength of all conversion and contact
effects described above,
whereas in~(\ref{gceqs}) the conversion mechanism involves a more traditional bilinear form.
In general, the matrix $\vec{g}=\{g_{ab}\}$ is not symmetric.
More importantly, it is expected to be sparse, as only similar grammar rules
pertaining to different languages will influence each other.
Furthermore, the couplings $g_{ab}$ connecting nodes pertaining to different languages,
i.e., representing contact between distinct languages, must be positive,
whereas those between nodes pertaining to the same language may take both signs.
Finally, the denominator $Z(N)$ is there to ensure that the sum rule
\beq
\sum_aN_a=N
\eeq
holds, where $N$ is the total number of verbs
in all languages of the family under consideration.

Introducing the reduced effective time
\beq
s=\int_1^N\frac{\dd N'}{Z(N')}
\label{sdef}
\eeq
brings the evolution equations~(\ref{ceqs}) to the form
\beq
\frac{\dd N_a}{\dd s}=\eta_a\left(N_a+\sum_bg_{ab}N_b\right).
\label{aeqs}
\eeq
These equations are autonomous,
in the sense that they no longer involve any explicit time dependence.
They can be recast in matrix form:
\beq
\frac{\dd N_a}{\dd s}=\sum_bM_{ab}N_b,
\label{ameqs}
\eeq
with
\beq
M_{ab}=\eta_a(\delta_{ab}+g_{ab}),
\eeq
where $\delta_{ab}$ is the Kronecker symbol.

We are chiefly interested in the late-time regime of the evolution
described by the contact network under consideration.
There, all group sizes grow asymptotically as
\beq
N_a\approx V_a\,\e^{\lambda s},
\eeq
where $\lambda$ is the largest eigenvalue of the constant dynamical matrix $\vec{M}=\{M_{ab}\}$,
whereas the amplitudes~$V_a$ are proportional to the components
of the associated right eigenvector, such that
\beq
\lambda V_a=\sum_bM_{ab}V_b.
\label{egv}
\eeq
The total number of verbs thus grows as
\beq
N\approx V\,\e^{\lambda s},
\quad
V=\sum_aV_a.
\eeq
We expect that $\lambda$ and the $V_a$ are positive in realistic circumstances,
even when the dynamical matrix~$\vec{M}$ is not symmetric.
These expectations have been confirmed by a range of numerical explorations.

The relative sizes of the groups in the late-time regime
are therefore dictated by the components~$V_a$
of the leading eigenvector of the dynamical matrix $\vec{M}$.
The spectral problem at hand presents a formal analogy with Anderson localisation
within the tight-binding formalism~\cite{rmp,lagendijk,anderson50}.
More precisely,
the dynamical matrix~$\vec{M}$ is analogous to the tight-binding Hamiltonian ${\cal H}$
describing the motion of a single electron in a random potential.
The largest eigenvalue~$\lambda$ is analogous to the ground-state energy~$E_0$ of the electron.
Finally, the components $V_a$ of the associated right eigenvector
are analogous to the components of the ground-state wavefunction of this one-body problem.
From a very general viewpoint,
the geometry of the underlying network and the distribution of the couplings
determine the nature (extended, localised, fractal, etc.) of the wavefunction.
The analogy of our problem with that of Anderson localisation
implies that differing network geometries and model parameters
will lead to a rich diversity of behaviour in
the asymptotic distribution of verb group sizes.

\subsection{The linear chain}

We first investigate the idealised situation where nodes form an infinite linear chain,
with asymmetric couplings between nearest neighbours.
In this context,~(\ref{egv}) reads
\beq
\lambda V_n=\eta_n(V_n+g_{n,n-1}V_{n-1}+g_{n,n+1}V_{n+1}),
\label{tbe}
\eeq
with obvious notations.
For specificity, we consider the case where the node at the origin is favoured,
in the sense that its attractiveness parameter is $\eta_0=1$,
whereas all other nodes have $\eta_n=q<1$.

Consider the pristine case where
all couplings are the same $(g_{n,n-1}=g_{n,n+1}=g$).
In this simple situation,
the analogy with the tight-binding model is as follows.
The favoured node at the origin acts as an attractive impurity,
where the wavefunction is expected to be largest.
It can be checked that the largest eigenvalue~$\lambda$
corresponds to a localised impurity state of the form
\beq
V_n=z^{\abs{n}},
\label{vnzn}
\eeq
which falls off exponentially with the distance $\abs{n}$ to the origin.
The eigenvalue $\lambda$ and the decay constant $z$ are determined by the two equations
\beq
\lambda=1+2gz=q\left(1+g\left(z+\frac{1}{z}\right)\right),
\eeq
hence
\beqa
z&=&\frac{2qg}{w+1-q},
\nonumber\\
\lambda&=&1+\frac{4qg^2}{w+1-q},
\eeqa
with the notation
\beq
w=\sqrt{(1-q)^2+4q(2-q)g^2}.
\eeq

In the more general case, where the couplings $g_{n,n-1}$ and $g_{n,n+1}$ are
arbitrary and modelled as, say, quenched random variables,
the above picture of a localised impurity state around the favoured node,
with an exponentially decaying wavefunction, remains qualitatively correct.
To get more quantitative results, we look at
the theory of fluctuations in one-dimensional Anderson localisation
(see~\cite{CT} and the references therein).
The latter theory predicts that the amplitude~$V_n$ is distributed log-normally
in the asymptotic limit.
More precisely, for a large distance $n$, the logarithmic ratio
\beq
L_n=-\ln\frac{V_n}{V_0}
\label{lndef}
\eeq
is approximately distributed according to the Gaussian law
\beq
f(L_n)\approx\frac{1}{\sqrt{2\pi\gamma_2n}}
\,\exp\left(-\frac{(L_n-\gamma_1n)^2}{2\gamma_2n}\right),
\label{pln}
\eeq
where $\gamma_1$ and $\gamma_2$ are the first two Lyapunov expo\-nents
of the problem, such that\footnote{The impurity state~(\ref{vnzn})
of the pristine case described above
fits within this scheme, with $\gamma_1=-\ln z$ and $\gamma_2\to0$.}
\beq
\mean{L_n}\approx\gamma_1n,
\quad
\var L_n\approx\gamma_2n.
\eeq

The probability of winning against the odds
for a node at a distance $n$ from the favoured one sitting at the origin is defined as
\beq
P_n=\prob{V_n>V_0}=\prob{L_n<0}.
\eeq
Using~(\ref{pln}), this becomes
\beq
P_n\approx\frac{1}{2}\erfc\left(\gamma_1\sqrt{\frac{n}{2\gamma_2}}\right).
\eeq
This expression falls off exponentially with distance $n$,
according to
\beq
P_n\sim\e^{-\mu n},
\label{pexp}
\eeq
with
\beq
\mu=\frac{\gamma_1^2}{2\gamma_2}.
\label{mu}
\eeq

In our context,
the above analysis suggests that the probability of finding an unlikely winner
(e.g.~an irregular grammatical form)
decays exponentially with the graph distance $n$ between that form
and the closest most regular (or otherwise favoured) form.
The illustration of our formalism in the idealised geometry of an infinite chain
will serve as a template for the analysis
in the next subsections, where we will formulate our problem in more realistic settings.

\subsection{A two-dimensional `toy' network}

\begin{figure}[!htbp]
\begin{center}
\includegraphics[angle=0,width=.7\linewidth,clip=true]{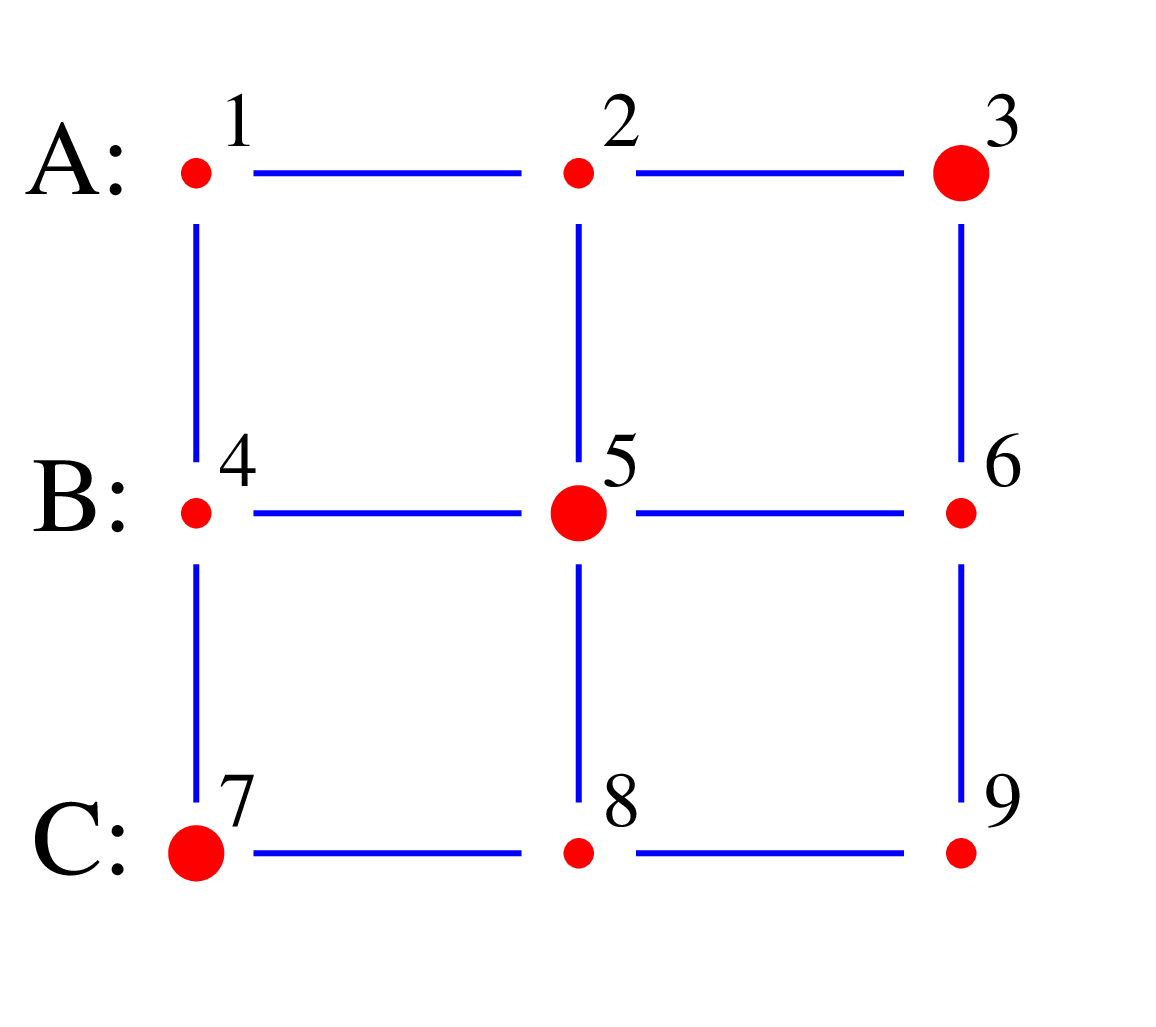}
\caption{
Schema of the toy network with 3 languages, denoted A, B and C,
and 3 verb groups in each language.
Large symbols: favoured nodes.
Small symbols: unfavoured nodes.
Blue lines: bonds carrying either symmetric or asymmetric couplings.}
\label{33net}
\end{center}
\end{figure}

We now look for winners against the odds in the setting of a $3\times3$ network
involving three related languages, denoted A, B and C (see Figure~\ref{33net}).
Individual nodes in a row correspond to verb groups in a given language;
we consider thus a total of three verb groups in each of the three languages.
This geometry is far more realistic than the previous one of an infinite linear chain,
even though it is not motivated by a specific example.
Three is indeed the right order of magnitude for the number
of verb groups, and more generally for competing grammatical forms.
It is also expected to be a good estimate of the number of closely related languages
with significant borrowing exchanges.
One may think of English, Dutch and German.

The conjugation rules of the verb groups (in different languages) which are
aligned vertically in a column are assumed to be very similar to each other.
A single verb group is favoured in each language (large symbols),
i.e., its attractiveness reads $\eta_a=1$, whereas the other two groups have
$\eta_a=q<1$ (small symbols).
The couplings (blue lines) are limited to nearest neighbours.
It is when the most favoured groups are chosen to be different in the three languages that
non-trivial behaviour emerges.

We study both symmetric and asymmetric isotropic random couplings.
In the symmetric case, the couplings $g_{ab}=g_{ba}$ along the 12 bonds
are independently drawn from the exponential law of parameter $\Delta$:
\beq
f_\Delta(g)=\frac{\e^{-g/\Delta}}{\Delta}.
\eeq
In the asymmetric case, both $g_{ab}$ and $g_{ba}$
are two independent positive random variables drawn from the above distribution,
so that there are altogether 24 random couplings.

We consider the fates of nodes 1 and~9,
which are furthest from the favoured nodes, and so are least likely to win.
For this purpose, we examine the logarithmic ratio (see~(\ref{lndef}))
\beq
L=-\ln\frac{V_1}{V_3}.
\label{ldef}
\eeq
The high symmetry of the network
implies that considering the alternative ratios $V_1/V_7$, $V_9/V_3$ and $V_9/V_7$
would yield statistically identical results.
A negative value of $L$ implies that node~1 wins against the odds;
in other words, we have $V_1>V_3$, despite mode 3 being the most attractive in language~A
(Note that there is no {\it direct} coupling between nodes 1 and 3).
The probability that node~1 wins against the odds therefore reads
\beq
P_\win=\prob{L<0}.
\label{pwindef}
\eeq

\begin{figure}[!htbp]
\begin{center}
\includegraphics[angle=0,width=1\linewidth,clip=true]{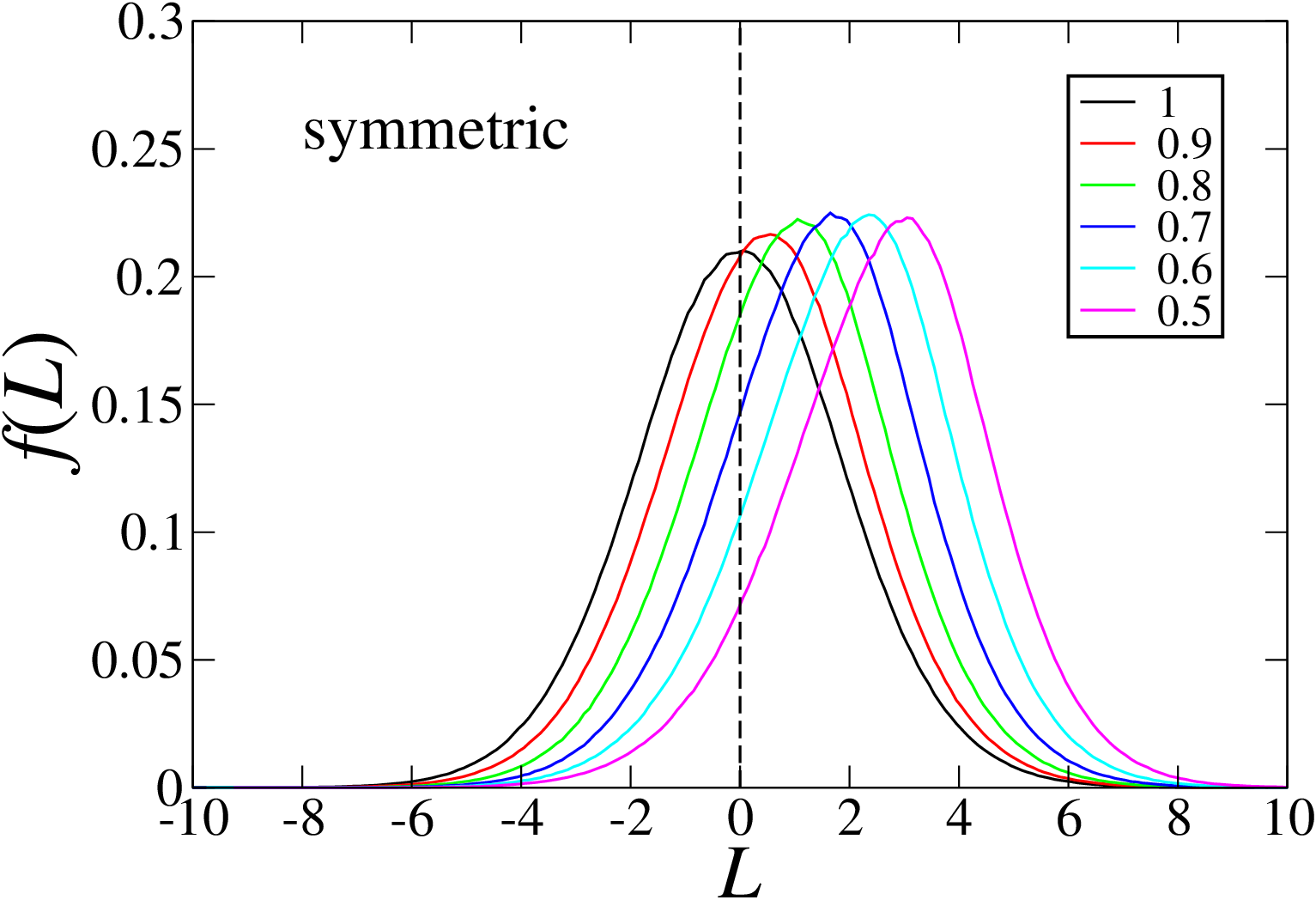}

\includegraphics[angle=0,width=1\linewidth,clip=true]{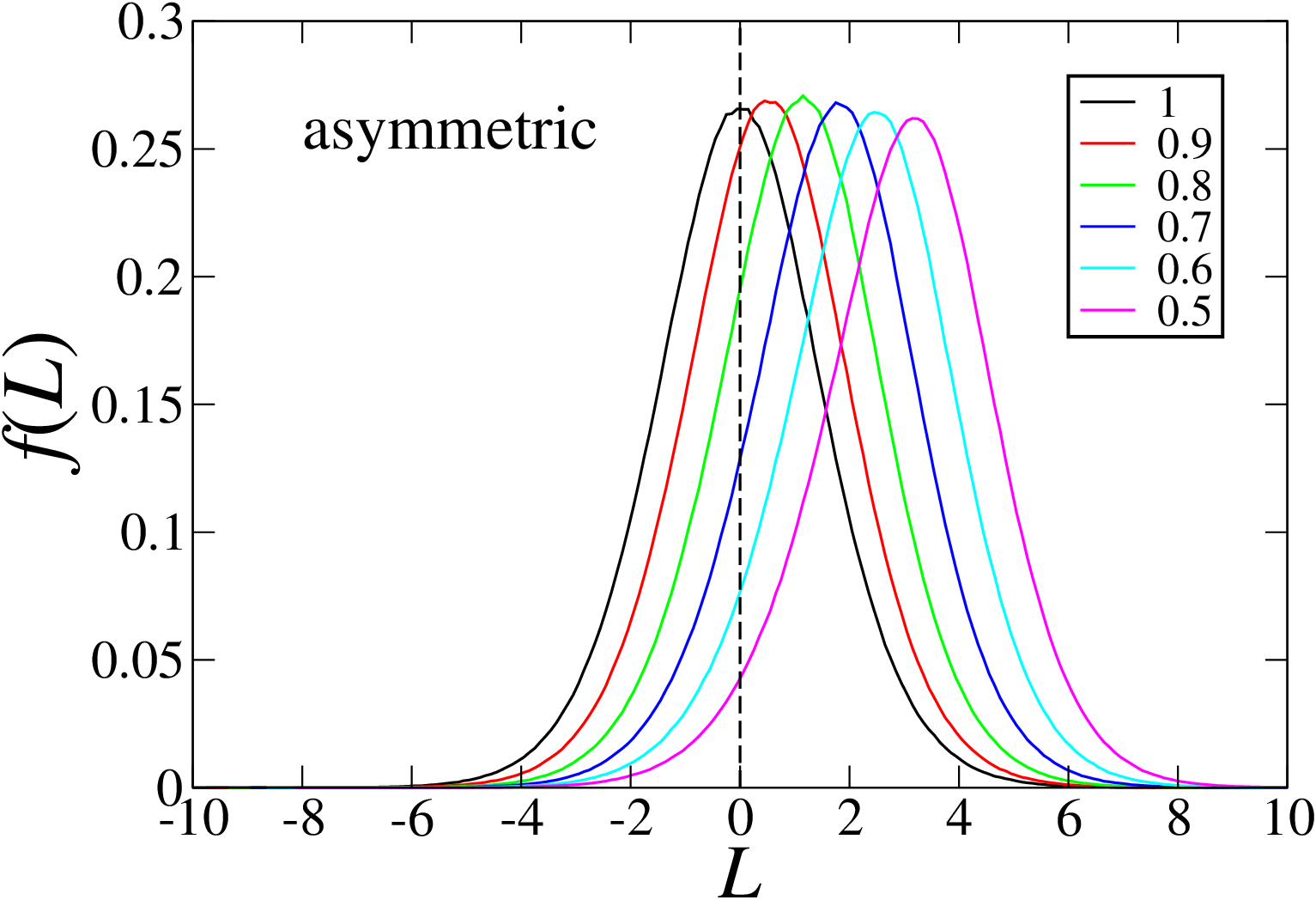}
\caption{
Probability distribution of the logarithmic ratio~$L$ (see~(\ref{ldef})),
for a fixed small coupling width $\Delta=0.1$
and several attractiveness ratios $q$ (see legend).
Vertical dashed lines: neutral value ($L=0$).
Upper panel: symmetric couplings.
Lower panel: asymmetric couplings.}
\label{toysas}
\end{center}
\end{figure}

\begin{figure}[!htbp]
\begin{center}
\includegraphics[angle=0,width=1\linewidth,clip=true]{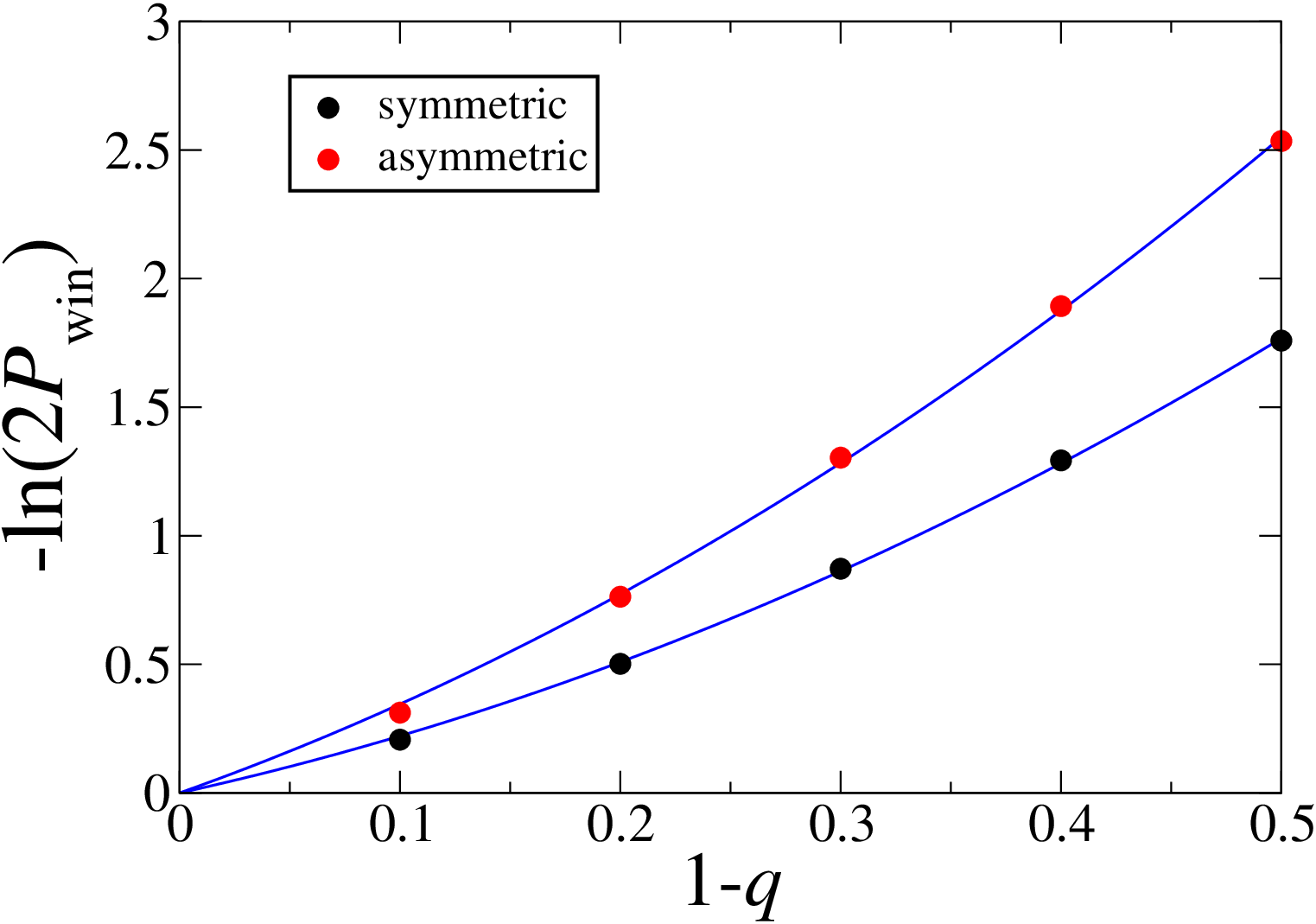}
\caption{
Plot of $-\ln(2P_\win)$ against the attractiveness contrast $1-q$,
where $P_\win$ is the probability that node~1 wins against the odds (see~(\ref{pwindef})).
Symbols: data extracted from those shown in Figure~\ref{toysas}
for symmetric and asymmetric couplings.
Blue curves: quadratic fits
suggesting that~$P_\win$ falls off faster than exponentially
as a function of the attractiveness contrast $1-q$.}
\label{pwin}
\end{center}
\end{figure}

Figure~\ref{toysas} shows plots of the probability distribution $f(L)$
of the logarithmic ratio $L$
for symmetric (top) and asymmetric (bottom) couplings,
with a fixed small coupling width $\Delta=0.1$
and several attractiveness ratios $q$ (see legend).
Data have been obtained by numerically solving the $9\times9$ eigenvalue equation~(\ref{egv})
for many independent draws of the random couplings.
Even though the network size is small,
the overall shape of the plotted distributions is close
to the asymptotic Gaussian profile~(\ref{pln}) obtained on the infinite chain.
The asymmetry of the couplings appears to play a rather minor role,
in the sense that both series of curves are rather similar to one another.
When $q=1$ (black curves), so that the attractiveness is the same throughout the network,
the distributions are symmetric around the neutral value $L=0$ (vertical dashed lines),
and $P_\win=1/2$, as expected.
As $q$ is decreased, the distributions shift progressively to the right,
with~$\mean{L}$ growing steadily with the difference $1-q$,
whereas their shapes remain roughly unchanged.
The portion of the curves
corresponding to negative values of $L$ shrinks accordingly,
indicating that winning against the odds becomes increasingly difficult
as the attractiveness contrast $1-q$ is increased.
This observation is made quantitative in Figure~\ref{pwin},
showing that the probability $P_\win$ of winning against the odds (see~(\ref{pwindef}))
falls off more rapidly than exponentially with the attractiveness contrast $1-q$.

The present network embedding,
with its explicit depiction of contact with neighbouring languages,
is more apposite than the setting of the infinite chain
for the problem of unlikely winners among grammatical forms.
However, the striking resemblance between
the distributions presented in Figure~\ref{toysas} for our rather small toy network
and the Gaussian profile~(\ref{pln}) for the infinite chain
is a strong indicator of the relevance of the theory of one-dimensional Anderson localisation
to the present problem.
This analogy enables us to reduce the problem of finding unlikely winners in
a dynamical system, where many agents are in simultaneous competition,
to one involving the exponential decay of $P_\win$ with distance from the nearest favoured node.
We will make good use of this simplification in Section~\ref{nets},
when the collective competition intrinsic to our problem is embedded on complex networks.

\subsection{Complex networks}
\label{nets}

A `map' of linguistic influences can be expected to have a complex topography,
comprising regions of strong linguistic contact (i.e., high connectivity),
as well as relatively isolated regions with weak or no linguistic contact.
It is the former that are relevant in the context of the question we ask: can
the emergence of irregular grammatical forms which survive against the odds
be attributed to the influence of `neighbouring' languages?
We therefore home in on what will be the most sophisticated,
as well as the most abstract, version of our model;
here, all grammar rules (governing verb conjugation, in this instance)
in a family of related languages are embedded
in a complex network (see e.g.~\cite{lnr,dm}).

We choose for definiteness the geometry of random regular graphs (see e.g.~\cite{janson,bol}).
These are randomly connected networks
where each node has the same prescribed degree $k\ge3$,
i.e., each node is connected to exactly $k$ other nodes.
The main qualitative features of the model, to be described below,
would remain essentially the same for other structureless models of complex networks,
where degrees $k$ have mild fluctuations around the mean degree $\mean{k}$.
Scale-free networks,
where node degrees $k$ obey a broad power-law distribution,
might however give rise to a different phenomenology.

Some of the connections involve two nodes in the same language,
while others embody contact with neighbouring languages;
as before, the bonds between nodes in the same language represent conversion,
while those connecting nodes belonging to different languages represent
interlingual contact.
We do not -- unlike for the toy network of the
previous subsection -- specify which is which; we focus instead
only on the nodes which win against the odds.
The distribution of attractiveness parameters is again assumed to be bimodal.
Some fraction~$\rho$ of the nodes are favoured,
and have attractiveness parameters $\eta_a=1$,
whereas all other nodes have the same smaller attractiveness $\eta_a=q$.
The microscopic distribution of couplings $g_{ab}$
entering the dynamical matrix is left unspecified.
We indeed rely on the key outcome of the analysis of the one-dimensional setting, i.e.,
the exponential falloff of the probability of winning against the odds
with distance~$n$ to the nearest favoured node (see~(\ref{pexp})).
For specificity, we assume a purely exponential decay law of the form
\beq
P_n=\e^{-\mu n}.
\label{pn}
\eeq
The parameter $\mu$ increases with the attractiveness contrast $1-q$,
in a way which depends on the distribution of the couplings $g_{ab}$, as suggested by~(\ref{mu}).
The greater the attractiveness contrast, therefore, the less will be the likelihood of
finding winners against the odds.

The central question we wish to address concerns the probability $P_\win$
that any given node (e.g., the origin~O) wins against the odds,
as a function of the model parameters $\rho$ and $\mu$.
This can be written as
\beq
P_\win=\sum_{n\ge1}f_nP_n,
\label{psum}
\eeq
where $f_n$ is the distribution of the distance $n$ of the origin~O to the nearest favoured node.
Note that the term $f_0=\rho$ does not enter the sum in~(\ref{psum}),
since this corresponds to the current node being favoured,
and therefore does not contribute to winners against the odds.

The distribution $f_n$ is evaluated as follows.
The number $M_n$ of nodes situated at distance at most~$n$ from the origin~O
reads
\beqa
M_n&=&1+k+k(k-1)+\cdots+k(k-1)^{n-1}
\nonumber\\
&=&\frac{k(k-1)^n-2}{k-2}.
\label{nn}
\eeqa
This result relies on
the property that a random regular graph is locally treelike,
so that cycles are rare, and therefore negligible.
In other words, local properties of random regular graphs coincide
with those of the infinite Cayley tree of degree $k$,
also known as the Bethe lattice in the physics literature.
The exponential growth law~(\ref{nn})
of the number of nodes with distance is an expression of the fact that the
fractal dimension of the network is formally infinite.
The diameter $n$ of a large network indeed grows logarithmically with its total mass $M$, as
\beq
n\approx\frac{\ln M}{\ln(k-1)}.
\eeq
This is a manifestation of the so-called {\it small-world effect}~\cite{lnr,dm}.

We now use the result~(\ref{nn}) to evaluate the probability $F_n$ that the distance
between~O and the nearest favoured node is larger than $n$.
This is identical to the probability that all the $M_n$ nodes
in the first $n$ shells around~O are unfavoured, so that $F_n=(1-\rho)^{M_n}$.
The distribution that we seek to evaluate is then nothing but the difference
\beq
f_n=F_{n-1}-F_n\qquad(n\ge1),
\eeq
i.e.,
\beq
f_n=(1-\rho)^{M_{n-1}}\left(1-(1-\rho)^{k(k-1)^{n-1}}\right).
\label{fn}
\eeq

The probability of winning against the odds is now readily obtained
by inserting~(\ref{pn}) and~(\ref{fn}) into~(\ref{psum}).

An interesting scaling regime takes place when the density $\rho$ of favoured nodes is small.
There, the typical distance to the nearest favoured node is large.
The distribution $f_n$ is peaked around a well-defined mean distance,
which grows logarithmically as
\beq
n^\star\approx\frac{\abs{\ln\rho}}{\ln(k-1)},
\eeq
with a bounded variance around this mean value.
Two distinct regimes emerge,
according to whether the sum entering~(\ref{psum}) is dominated
by the first few values of $n$ or by $n\approx n^\star$.
These regimes are defined by comparing two inverse lengths,
viz.~$\mu$, characterising the exponential decay law~(\ref{pn})
of the winning probability~$P_n$,
and $\ln(k-1)$, characterising the exponential proliferation~(\ref{nn})
of nodes around a given node.

\begin{itemize}

\item
For $\mu>\ln(k-1)$, the distant-dependent probabilities $P_n$
fall off fast enough that the sum in~(\ref{psum}) is dominated by finite values of $n$,
i.e., $n\ll n^\star$.
Winners against the odds are actually not too far from being likely winners,
since the nearest favoured node is nearby.
For small $\rho$, we have
\beq
f_n\approx k(k-1)^{n-1}\rho,
\eeq
and therefore
\beqa
P_\win&\approx&\sum_{n\ge1}k(k-1)^{n-1}\e^{-\mu n}\rho
\nonumber\\
&\approx&\frac{k\rho}{\e^\mu-(k-1)}
\label{phigh}
\eeqa
starts increasing linearly in $\rho$.

\item
For $\mu<\ln(k-1)$, the distance-dependent probabilities $P_n$
fall off slowly enough that the sum in~(\ref{psum}) is dominated by $n\approx n^\star$.
Winners in this case are genuinely against the odds,
since the nearest favoured node is quite distant.
We have
\beq
P_\win\sim P_{n^\star}\sim\rho^\alpha,
\label{plow}
\eeq
where the growth exponent $\alpha$ depends linearly on $\mu$, according to
\beq
\alpha=\frac{\mu}{\ln(k-1)}.
\label{alow}
\eeq

\item
In the borderline case where $\mu=\ln(k-1)$, the summand
\beq
f_nP_n\approx\frac{k\rho}{k-1}
\eeq
is nearly flat up to $n^\star$.
We thus obtain
\beq
P_\win\approx\frac{k\rho\abs{\ln\rho}}{(k-1)\ln(k-1)}.
\label{pmarg}
\eeq

\end{itemize}

We pause briefly to consider the significance of the parameter $\alpha$ (see~(\ref{alow})).
A similar ratio characterised survivors in an earlier model
of competitive dynamics on networks~\cite{us2}, where the
probability of survival of a node depended on the ratio of its mass to its
average degree distribution; the `heavier' the node, and the less connected it
was to others, the better its chances of survival.
In the present case, the role of the mass in~\cite{us2} is played by the
the attractiveness contrast $(1-q)$ (recall that ${\mu}\sim(1-q)$), with the
parameter $\ln(k-1)$ representing the effect of the (constant) degree distribution.
This analogy puts our work into a more general context: {\it the more
attractive a grammar rule is, and the less connected it is to direct
competitors, the more it is likely to persist}.\footnote{The important difference
is that {\it all} survivors were considered in~\cite{us2}, whereas here only
the subset of survivors against the odds is considered.
The same reasoning though clearly applies to both.}

\begin{figure}[!htbp]
\begin{center}
\includegraphics[angle=0,width=1.\linewidth,clip=true]{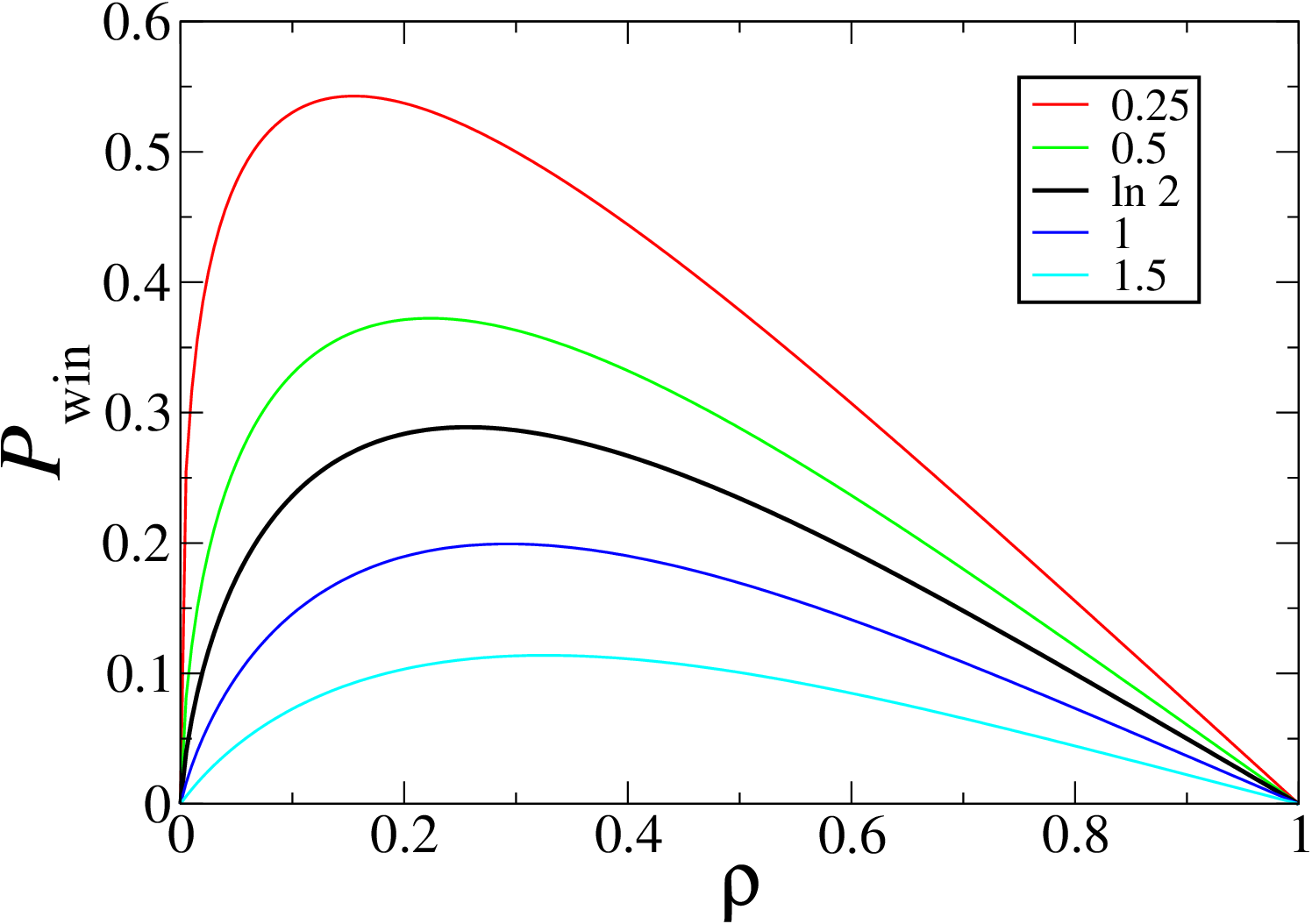}
\caption{
Probability $P_\win$ of winning against the odds for a typical node of a random regular graph,
against the density~$\rho$ of favoured nodes,
for $k=3$ and several values of $\mu$ (see legend).}
\label{prrn}
\end{center}
\end{figure}

Figure~\ref{prrn} shows plots of the probability $P_\win$
of winning against the odds against the density $\rho$ of favoured nodes,
for $k=3$ and several values of the parameter $\mu$ (see legend).
The initial rise of $P_\win$ at small $\rho$ is faster than linear for the two
upper curves ($\mu<\ln 2$) (see~(\ref{plow}))
and linear for the two lower curves ($\mu>\ln 2$) (see~(\ref{phigh})).
The borderline case ($\mu=\ln 2$) (see~(\ref{pmarg})) is also shown (thick black curve).
In the other limiting situation $(\rho\to1)$, only
\beq
f_1=(1-\rho)(1-(1-\rho)^k)
\label{f1res}
\eeq
scales linearly with the difference $1-\rho$, so that
\beq
P_\win\approx\e^{-\mu}(1-\rho).
\eeq

For any value of the parameter $\mu$,
the probability $P_\win$ vanishes both for $\rho\to0$ and for $\rho\to1$.
This must clearly be the case: the complete absence of favoured sites
will not cause winners against the odds to be generated at all, while if nearly all sites are
favoured, winners will be very much with the odds, and not against them.
We notice also that $P_\win$ is overall larger for $\mu<\ln(k-1)$ than for $\mu>\ln(k-1)$.
This ties up with the arguments we were making above: we would expect that
winners would be more numerous when there are many unfavoured nodes
under the umbrella, so to speak, of a favoured site (see~(\ref{plow})) than in
the opposite situation (see~(\ref{phigh})).

\begin{figure}[!htbp]
\begin{center}
\includegraphics[angle=0,width=1.\linewidth,clip=true]{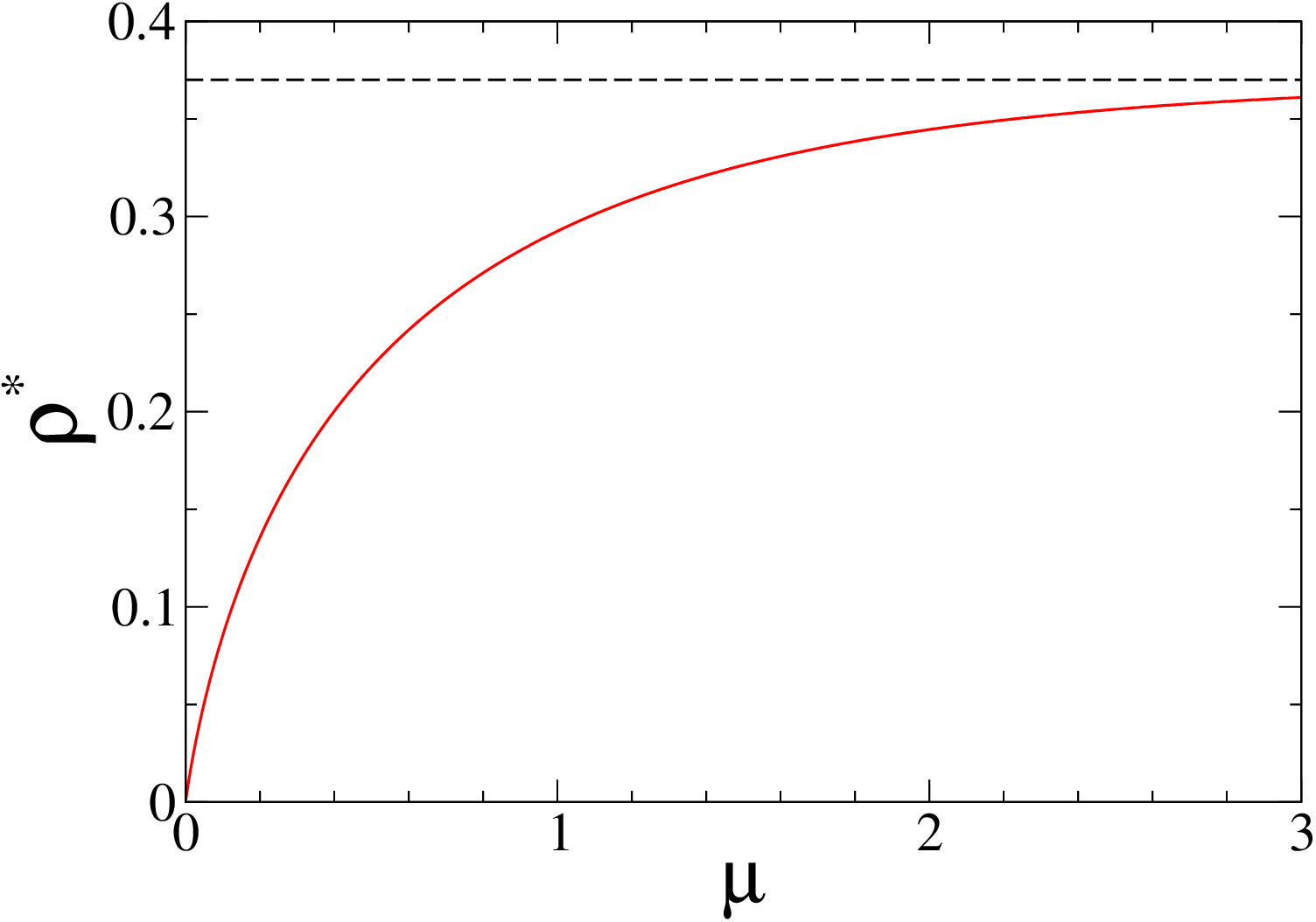}
\caption{
Optimal density $\rho^\star$ of favoured nodes,
such that the probability $P_\win$ is maximal, against the parameter $\mu$ for $k=3$.
Dashed horizontal line: limiting value
$\rho_\infty=1-4^{-1/3}\approx0.370039$.}
\label{rhostar}
\end{center}
\end{figure}

The plots in Figure~\ref{prrn} also clearly manifest the presence of an
optimal density $\rho^\star$ of favoured nodes,
where $P_\win$ reaches its maximal value $P_\win^\star$.
Figure~\ref{rhostar} shows a plot of this optimal density against $\mu$ for $k=3$.
At small $\mu$, $\rho^\star$ starts growing linearly.
At large $\mu$, we have $P_\win\approx\e^{-\mu}f_1$ (see~(\ref{f1res})),
so that~$\rho^\star$ saturates to the limiting value
\beq
\rho_\infty=1-(k+1)^{-1/k}.
\eeq
(The precise values of $P_\win^\star$ are however less informative,
as they depend on the assumed exact exponential form with unit
amplitude~(\ref{pn}) of $P_n$.)
The interpretation of these plots is as follows: when $\mu$ is small, one does not need
a high density of favoured sites to achieve $P_\win^\star$; the unfavoured
sites themselves are close enough to being winners.
When, however, $\mu$ is large, we need a far higher density of favoured sites to have
a sphere of influence strong enough to attain $P_\win^\star$.
The optimal density $\rho^\star$ is therefore an increasing function of
$\mu$ until it saturates to $\rho_\infty$.

\begin{figure}[!htbp]
\begin{center}
\includegraphics[angle=0,width=1.\linewidth,clip=true]{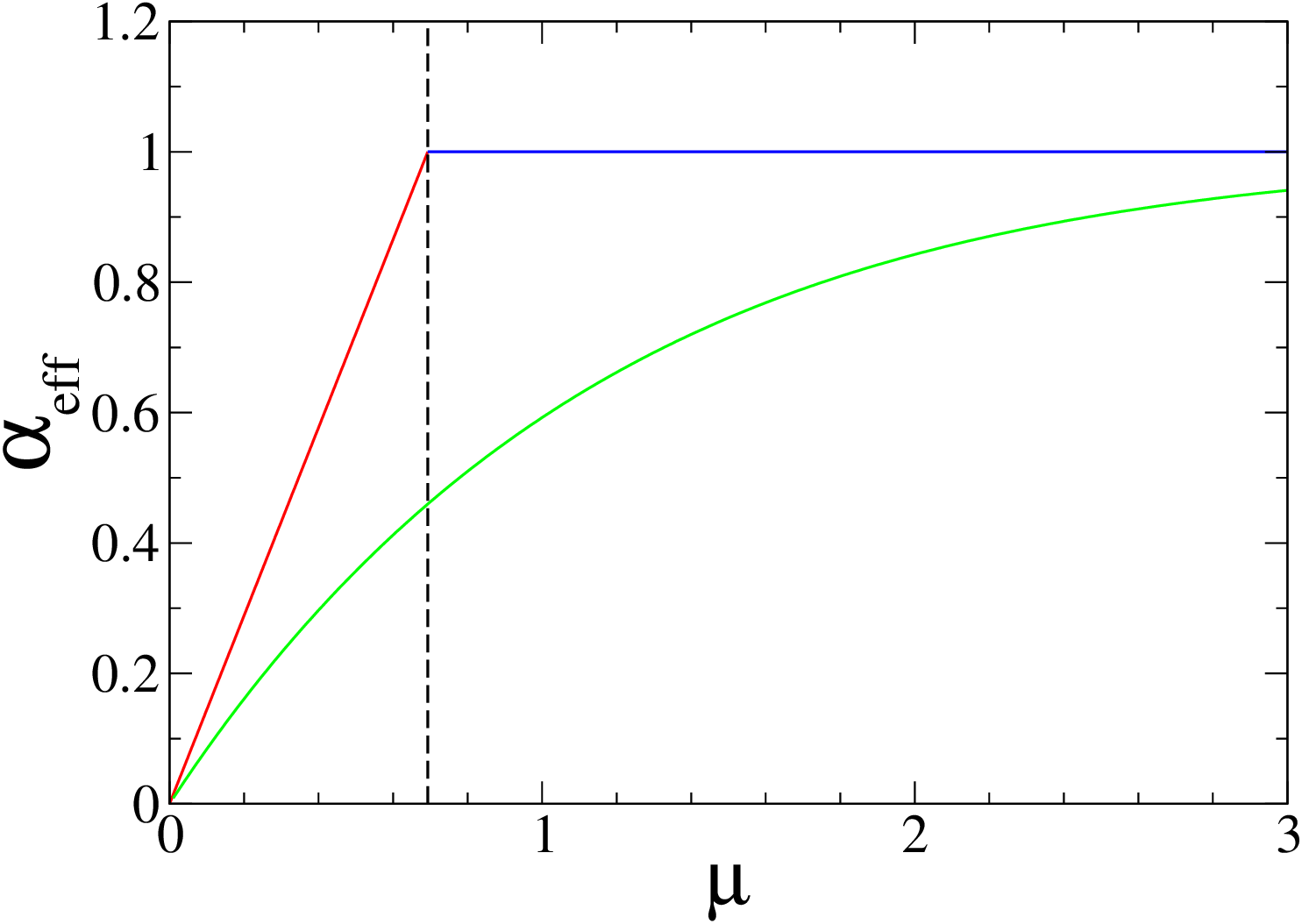}
\caption{
Green curve:
effective exponent $\alpha_\eff$ entering the approximate formula~(\ref{pfit}),
against $\mu$ for $k=3$.
Red line: exact growth exponent~(\ref{alow}) for $\mu<\ln 2$.
Blue horizontal line: exact unit exponent of the linear law~(\ref{phigh}) for $\mu>\ln 2$.
Vertical dashed line: borderline value $\mu=\ln 2$.}
\label{alpha}
\end{center}
\end{figure}

It turns out that the overall dependence of $P_\win$
on the density $\rho$ of favoured nodes
is always rather accurately represented by the phenomenological formula
\beq
P_\win\approx\rho^{\alpha_\eff}-\rho.
\label{pfit}
\eeq
Figure~\ref{alpha} shows a plot of the effective exponent $\alpha_\eff$,
obtained by means of a nonlinear fit, against~$\mu$ for $k=3$.
This effective exponent (green curve) interpolates smoothly between
the two exact growth exponents in the low-density scaling regime (red and blue lines).

We now summarise and interpret the above results.
The linguistic map in a region of strong linguistic contact, alluded to
at the beginning of this section, is constructed by embedding our model of
competing grammar rules on a sparse complex network,
here chosen to be the Bethe lattice.\footnote{We mention for completeness that the present work
has only little to do with the theory of Anderson localisation on the Bethe lattice~\cite{mf},
which plays a major role in recent work on many-body localisation
(see e.g.~\cite{dla,kai,tik}).
There, the Bethe lattice arises as a template for the Fock space of a quantum many-body problem.}
The mechanisms of competition involve both linguistic contact between different languages
and regularisation within a given language, which act against each other.
Our analysis is based on an effective description, via analogies with Anderson localisation,
of the eigenvector associated with
the largest eigenvalue $\lambda$ of the dynamical matrix~$\vec{M}$ on this lattice, and
involves only its exponential decay with distance from the set of favoured nodes.
We have discovered the existence of a low-density scaling regime, where
irregular forms vastly outnumber regular ones and two kinds of unlikely winners emerge:
the first correspond to grammatical forms which are linguistically close to a
favoured regular form,
while the second correspond to those which are linguistically rather distant from it.

These results provide a natural framework
within which the emergence of `stuck' can be explained.
According to Ringe~\cite{longringe},
the competing past participles of the verb `stick' were `sticked', `stuck', `stoke' and
`stoked', and the application of the tolerance principle, which takes a
one-body view of the problem, did not predict the winner.
In our formalism, {\it all these coexisting grammatical
forms compete with each other in parallel in a dynamical model}.
In the relevant dynamical regime,
where irregular forms far outnumber the only `regular' competitor `sticked'
($\rho$ small), the scenario of~(\ref{plow}) applies,
and a deeply irregular form such as `stuck',
linguistically very distant from the regular form `sticked', emerges as a winner against the odds.

\section{Discussion}
\label{disc}

We have, in the above, used statistical physics methodologies
to model the evolution of grammatical rules.
Starting with a very general approach,
we have provided a useful way of classifying rules and exceptions in typical situations.
A major result of this static approach is that
exceptions are so called for a good reason: i.e., that they occur rarely,
their number growing either logarithmically or as a subextensive power law in the number of items.
Our work quantifies a well-known example from linguistics, viz.~the paucity of
verb groups in most world languages, by demonstrating that the birth of
a new grammatical rule (or a new verb group) is a very rare event
(see the logarithmic laws~(\ref{log1}),~(\ref{klog}),~(\ref{mlog})).
All of the above is in stark contrast to the high threshold predicted by the
tolerance principle, which has a nearly extensive
(i.e., nearly maximal) growth law for the number of allowed exceptions.

The dynamical models we have presented later in the paper have corroborated
these insights; we have there focused on morphological issues
in verb conjugation, and their evolution.
The two main mechanisms we have included in our models are first,
the conversion of forms from irregular to regular within a given language, and second,
the influence of `neighbouring' languages with which it is in prolonged contact.
We have built up our models progressively, adding these ingredients one at a time
to observe their full consequences.
Our first dynamical model of Section~\ref{growth} involves only the
growth of the lexicon and of grammar rules in a single language in the first stage of its evolution.
In Section~\ref{gandc}, we have added a conversion mechanism, which describes the
rather universal tendency towards regularisation observed by linguists~\cite{lieberman}.
Finally, in Section~\ref{comparative}, we have completed the picture by adding
linguistic change via prolonged contact between similar languages, and argued that
this mechanism might well be responsible for the
introduction of novel irregular forms into a given language.

We have strongly emphasised
the appearance and persistence of winners or survivors against the odds
in all our dynamical models of linguistic evolution;
these unlikely winners are precisely the irregular linguistic
forms which persist in the face of several competitors, irregular as well as regular.
All the models we present include this essential ingredient
of collective and simultaneous competition.
In Sections~\ref{gandc} and~\ref{comparative},
we have quantified the probability of occurrence of these winners against the odds,
as a function of model parameters in a variety of situations.
Finally, we have constructed a model linguistic map in Section~\ref{nets}
incorporating both
intra-language regularisation and inter-language contact, and shown that
forms which are linguistically very far from favoured can indeed emerge, and
persist; in particular, this scenario allows us
to explain the unlikely persistence of the deeply irregular grammatical forms
mentioned in~\cite{longringe}.
It seems very likely that
such persistence occurs when the influence of interlingual contact exceeds
that of intralingual regularisation.

The quantitative proof of the above contention would entail the formulation by quantitative linguists
of realistic network models of language neighbourhoods.
These would require a full knowledge of
linguistically appropriate distributions of conversion and contact interactions for
closely linked language groups (see~Section~\ref{nets}).
Such distributions might, for example, be obtained from numerical experiments on the phylogenesis
of selected grammatical forms (past participles in our case) in the languages concerned.
It would be most interesting to see if our model predictions are verified, i.e.,
if, in the case where contact exceeds conversion, unlikely irregular forms survive.

In summary, the main conclusions of our work are as follows.
First, attractive grammar rules survive best in the absence of strong competitors
within or outside the language concerned;
second, there is an optimal density of favoured rules
which maximises the probability of less attractive rules winning {\it against the odds};
third, this optimal density decreases in proportion to the difference
in attractiveness between favoured and unfavoured rules;
and finally, despite overall tendencies towards regularisation,
irregular forms may persist in a given language because of their strong similarities
with sufficiently attractive forms in other, closely related, languages.

\subsubsection*{Acknowledgments}

\small
We acknowledge with thanks discussions and ex\-changes with Aur\'elia Elalouf,
Guillaume Jacques and Mattis List,
which introduced us to the rich fields of historical linguistics and contact linguistics.
AM warmly thanks the Leverhulme Trust for the
Visiting Professorship that funded part of this research,
as well as the Faculty of Linguistics, Philosophy and Phonetics, Oxford
and the Institut de Physique Th\'eorique, Saclay, for their hospitality.

\subsubsection*{Author contribution statement}

\small
Both authors contributed equally to the present work,
were equally involved in the preparation of the manuscript,
and have read and approved the final manuscript.

\subsubsection*{Data availability statement}

\small
Data sharing not applicable to this article as no datasets were generated or analysed
during the current study.

\bibliography{paper.bib}

\end{document}